\documentclass{aa}
\usepackage{graphicx}
\usepackage{color}

\begin{document}

\title{Extreme Superluminal Motion in the Curved Jet of \object{PKS\, 1502+106}}

\author{ T.~An      \inst{1,2}
        \and X.Y. Hong  \inst{1,2}
        \and T. Venturi \inst{3}
        \and D.R. Jiang \inst{1,2}
    \and W.H. Wang  \inst{1,2}
         }%

\institute{Shanghai Astronomical Observatory, Chinese Academy of
Sciences, Shanghai 200030, China \and National Astronomical
Observatories, Chinese Academy of Sciences, Beijing 100012, China
\and Istituto di Radioastronomia del CNR, Via Gobetti 101, 40129
Bologna, Italy }

\offprints{T.~An, antao@center.shao.ac.cn}

\date{Received $<$date$>$ / Accepted $<$date$>$}

\abstract{%
In this paper we present a multifrequency and multiepoch study of
\object{PKS\, 1502+106} at radio frequencies. The analysis is
based on an EVN (European VLBI Network) dataset at 5 GHz and
archive VLBA (Very Long Baseline Array) datasets at 2.3, 8.3, 24.4
and 43.1 GHz over a period of 8 years. The source is characterized
by a multi--component one--sided jet at all epochs. The
high--resolution images at 5, 8.3, 24.4 and 43.1 GHz show a curved
jet morphology in the source. The radio core brightness
temperature approaches the equipartition limit. Superluminal
motions of $37.3\pm9.3\;c$, $22.0\pm15.5\;c$, $10.5\pm2.6\;c$ and
$27.9\pm7.0\;c$  are measured in four distinct components of the
jet. Our analysis supports the idea that the relativistic jet in
\object{PKS\,1502+106} is characterised by extreme beaming and
that its radio properties are similar to those of $\gamma$--ray
loud sources.
\keywords{galaxies: jets -- galaxies: quasars: general --
galaxies: quasars: individual: \object{PKS\, 1502+106}}}

\maketitle \markboth{T.~An, X.Y.~Hong \& T.~Venturi et
al.}{Extreme Superluminal Motion in the Curved Jet of
\object{PKS\, 1502+106}}

\section{Introduction}

\begin{table*}
\centering \caption{Logs of the observations}
\begin{tabular}{ccclc}
\hline

Epoch &Freq  &BW &Array and Available Telescopes$^{a}$&$D_{uv}^{b}$  \\
      &(GHz) &(MHz)    &                           &(km)          \\
  (1) & (2)  & (3)     & (4)                       &(5)            \\\hline

1994.52 &2.3 & 16      &VLBA(All 10)     & 8600  \\

        &8.3 & 16      &VLBA(All 10)     & 8600  \\

1997.85 &5.0 & 28      & EVN(Ef Sh Jb Ht Mc Nt On Tr Ur Wb) & $\sim$10000 \\

1998.11 &2.3 & 32      &VLBA(BR FD MK OV PT)  GC KK      &             \\

        &    &         &VLBA(BR FD HN KP LA NL OV PT SC) GC WF GN &5600  \\

        &8.3 & 32      &VLBA(BR FD HN KP LA NL OV PT SC) GC WF GN &5600  \\

2002.37 &24.4& 32      &VLBA(BR FD HN KP LA MK NL OV SC)     & 8600   \\

        &43.1& 32      &VLBA(BR FD HN LA MK NL OV SC) & 8600  \\

2002.65 &24.4& 30      &VLBA(All 10)     & 8600  \\

        &43.1& 32      &VLBA(BR FD HN KP LA MK NL OV PT) & 6600 \\

2002.99 &24.4& 64      &VLBA(FD HN KP LA MK NL OV PT SC) & 8600 \\

        &43.1& 64      &VLBA(FD HN KP LA MK NL OV PT SC) & 8600 \\

\hline\end{tabular} \label{obs}
\\[0.3cm]
\raggedright $^{a}$ EVN telescope codes: Ef: Effelsberg, Sh:
Shanghai, Ht: HartRAO, Jb: Jodrell, Mc: Medicina, Nt: Noto, On:
Onsala, Tr: Torun, Ur: Urumqi, Wb: Westerbork Array; the VLBA
observations at epoch 1998.11 are performed using subarrays made
up of the VLBA antennas together with 4 geodetic antennas: GC
(Fairbanks, AK USA), WF (Westford, MA USA), GN (Green Bank, WV
USA) and KK (Kokee Park, HI USA);\\
$^{b}$ the longest baseline of the array, in unit of kilometer.
\end{table*}

\begin{table*}
\centering \caption{Parameters of the images}
\begin{tabular}{lccllccc}
\hline

Figure& Epoch & Freq  & Real Beam$^{a}$ & Restored
Beam$^{b}$ &S$_{Peak}~^{c}$ & rms$^{d}$ &Contours\\
      &       & (GHz) &Maj$\times$Min(mas),P.A.($\degr$)&
      Maj$\times$Min(mas),P.A.($\degr$)&(Jy/b) & (mJy/b)&(mJy/b)\\
(1) & (2) & (3) & (4) &(5) & (6) & (7) &(8) \\\hline
Fig.\ref{fig1}a&1994.52 &2.3 & 7.18$\times$3.78,
$-1.67$&7.18$\times$3.78, 0& 1.81&1.1 & 3.5$\times$(-1,1,2,...,256)\\

Fig.\ref{fig1}b&1998.11 &2.3 & 5.94$\times$4.54, 13.7
&7.18$\times$3.78, 0& 1.23&0.9 & 2.8$\times$(-1,1,2,...,256)\\

Fig.\ref{fig2}a&1997.85 &5.0 & 1.36$\times$1.21,  64.7
&1.25$\times$1.25, 0& 0.80&0.7 & 3.0$\times$(-1,1,2,...,256)\\

Fig.\ref{fig2}b&1994.52 &8.3 & 1.95$\times$1.03,
$-1.19$&1.25$\times$1.25, 0& 1.37&0.9 & 2.8$\times$(-1,1,2,...,256)\\

Fig.\ref{fig2}c&1998.11 &8.3 & 1.62$\times$1.23, 16.7
&1.25$\times$1.25, 0& 0.68&0.6 & 2.0$\times$(-1,1,2,...,256)\\

Fig.\ref{fig3}a&2002.37 &24.4& 0.64$\times$0.28,
$-$1.08&0.64$\times$0.28, 0& 1.26&0.8 & 2.1$\times$(-1,1,2,...,256)\\

Fig.\ref{fig3}b&2002.65 &24.4& 0.70$\times$0.32,
$-$3.77&0.64$\times$0.28, 0& 0.73&0.9 & 2.8$\times$(-1,1,2,...,256)\\

Fig.\ref{fig3}c&2002.99 &24.4& 0.76$\times$0.28, $-$6.0
&0.64$\times$0.28, 0& 0.95&1.0 & 3.0$\times$(-1,1,2,...,128)\\

Fig.\ref{fig4}a&2002.37 &43.1& 0.37$\times$0.16,
$-$3.35&0.37$\times$0.16, 0& 0.82&1.0 & 3.0$\times$(-1,1,2,...,256)\\

Fig.\ref{fig4}b&2002.65 &43.1& 0.66$\times$0.24,
$-$20.3&0.37$\times$0.16, 0& 0.50&1.2 & 3.7$\times$(-1,1,2,...,128)\\

Fig.\ref{fig4}c&2002.99 &43.1& 0.44$\times$0.16,
$-$6.41&0.37$\times$0.16, 0& 0.59&1.0 & 3.2$\times$(-1,1,2,...,128)\\
\hline
\end{tabular}\label{image}
\\[0.3cm]
\raggedright $^a$ the beam size directly measured from the
visibilities (FWHM);\\
$^b$ the restored beam shown in the images (FWHM);\\
$^c$ the peak brightness in the image; \\
$^d$ the off--source rms noise level in the images.
\end{table*}

One of the most significant observational results of extragalactic
$\gamma$--ray active galactic nuclei (AGNs) is that all
EGRET--identified objects are radio--loud sources (Mattox et al.
\cite{Mattox}). Relativistic beaming in the jet is used to explain
the EGRET identification in radio--loud AGNs. The
EGRET--identified sources have on average much faster apparent
superluminal motions than the general population of radio--loud
sources (Jorstad et al. \cite{Jorstad}). From a statistical
analysis of $\Delta PA$ (position angle differences between
parsec-- and kiloparsec--scale structures) of EGRET--identified
AGNs, Hong et al. (\cite{Hong}) concluded that $\gamma$--ray loud
radio quasars typically show aligned morphologies on parsec and
kiloparsec scales. It is still a matter of debate if the
$\gamma$--ray emission in AGNs is related to higher beaming in
these sources.

The radio--loud active galactic nucleus (AGN)
\object{PKS\,1502+106} (4C 10.39, OR103), $z=1.833$ (Fomalont et
al. \cite{Fomalont}), is a highly polarized quasar (Tabara \&
Inoue \cite{Tabara}). A high and variable degree of polarisation
in the optical band is reported by Impey \& Tapia (1988). It is
known to be active and variable at radio, optical and X--ray
wavelengths. In particular, the source exhibits intensity
variations by a factor of 3 $\sim$ 5 on timescales from weeks to
months in the radio band; it shows intensity variations in the
optical band with $m_{\nu}$ ranging from 19.5 to 18.6, and in the
X--ray band by a factor $\geq 2$ at 1~keV (George et al.
\cite{George} and references therein).

\object{PKS\, 1502+106} exhibits a 'core--jet--lobe' structure at
radio wavelengths. A Very Large Array (VLA) image at 1.64 GHz
(Murphy, Browne \& Perley \cite{Murphy}) shows that a continuous
jet extends to the southeast and leads to a lobe located $\sim$~7
arcsecond from the core. Very Long Baseline Interferometry (VLBI)
observations (Fey, Clegg \& Fomalont \cite{Fey}; Fomalont et al.
\cite{Fomalont}; Zensus et al. \cite{Zensus}) show a well--defined
jet starting to the southeast and sharply bending to the east at a
distance of 3--4 mas from the core.

Our interest in \object{PKS\,1502+106} is related to the
misalignment of the pc-- and kpc--scale radio structure in AGNs
and its relation to the $\gamma$--ray emission. Fichtel et al.
(\cite{Fichtel}) reported an EGRET flux density upper limit in PKS
1502+106 of $7 \times 10^{-8}$ photons cm$^{-2}$ s$^{-1}$ from the
Phase I results. However, the source was not detected in the
following EGRET observations (Thompson et al. \cite{Thompson};
Hartman et al. \cite{Hartman}). To study the relation between the
$\Delta PA$ and $\gamma$--ray emission, we observed a sample of
$\gamma$--ray blazars with the EVN at 5 GHz in Nov. 1997, in which
PKS 1502+106 was observed as a gamma--ray source candidate (Hong
et al. \cite{Hong04}).

In the next section we present the EVN observation of
\object{PKS\,1502+106} at 5 GHz, and the data reduction. The
analysis of the 5 GHz EVN image is performed in Sect. 3, where we
also re--analyzed five epochs of Very Long Baseline Array (VLBA)
datasets at 2.3, 8.3, 24.4 and 43.1 GHz obtained from the public
archive of Radio Reference Frame Database (RRFID) for a consistent
study with our data. To investigate the relativistic beaming in
this source, we estimate the core physical parameters and measure
the jet proper motions in Sect. 4. Conclusions and summary are
given in Sect. 5. $H_0$= 65 km\, s$^{-1}$Mpc$^{-1}$ and $q_0$=0.5
are used in this paper.

%

\begin{figure}\centering
\resizebox{6.5cm}{!}{\includegraphics{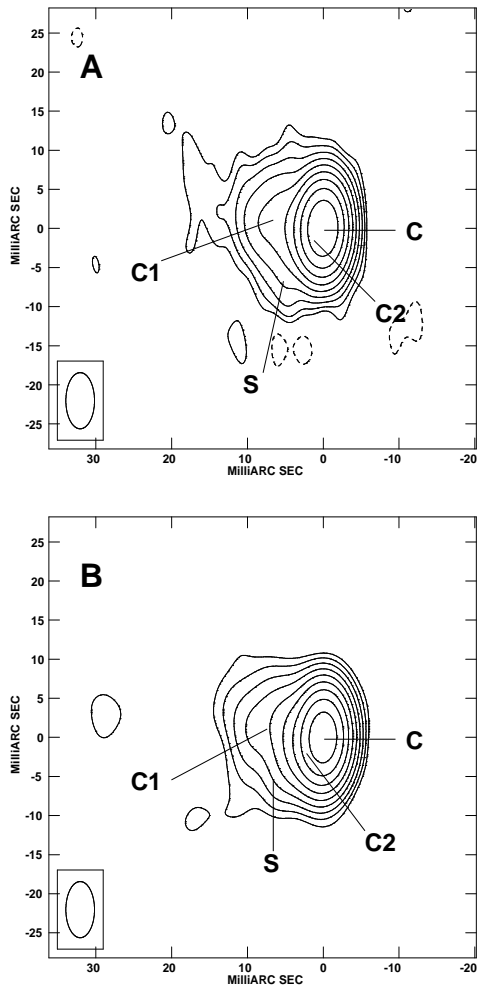}} \vspace{-1mm}
\caption{VLBA images of \object{PKS\, 1502+106} at 2.3 GHz. A:
epoch 1994.52; B: epoch 1998.11.} \label{fig1}
\end{figure}
%
%
\begin{figure}\centering
\resizebox{6.5cm}{!}{\includegraphics{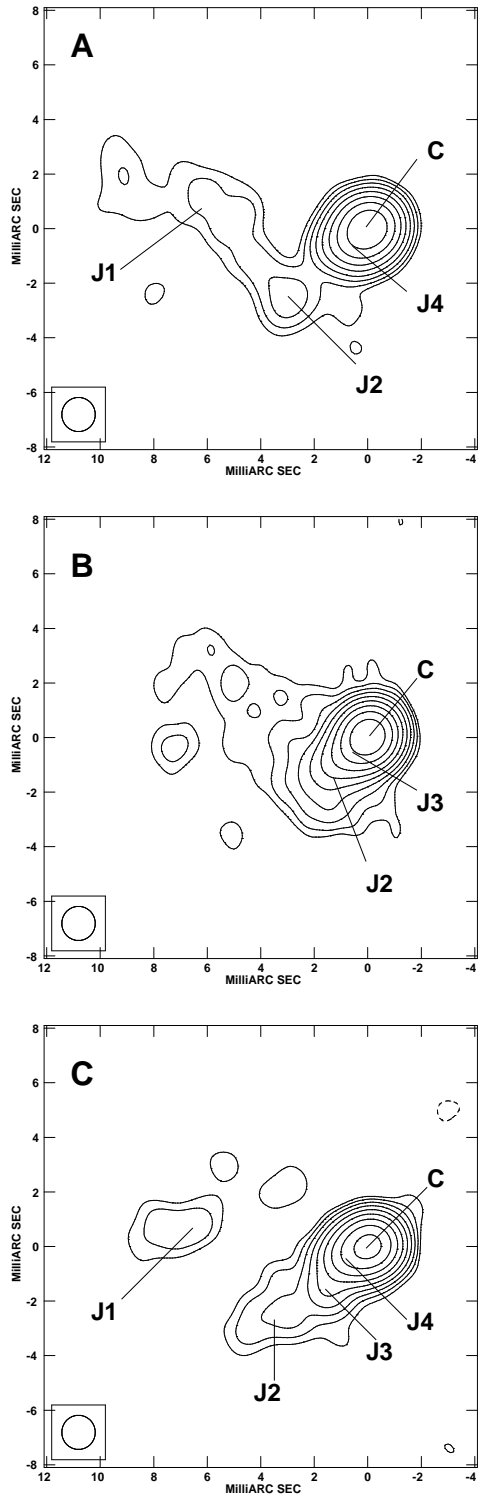}}\vspace{-1mm}
\caption{VLBI images of \object{PKS\, 1502+106}. A: EVN image at 5
GHz at epoch 1997.85; B: VLBA image of at 8.3 GHz at epoch
1994.52; C: VLBA image at 8.3 GHz at epoch 1998.11. } \label{fig2}
\end{figure}
%
\begin{figure}\centering \vspace{3mm}
\resizebox{6.5cm}{!}{\includegraphics{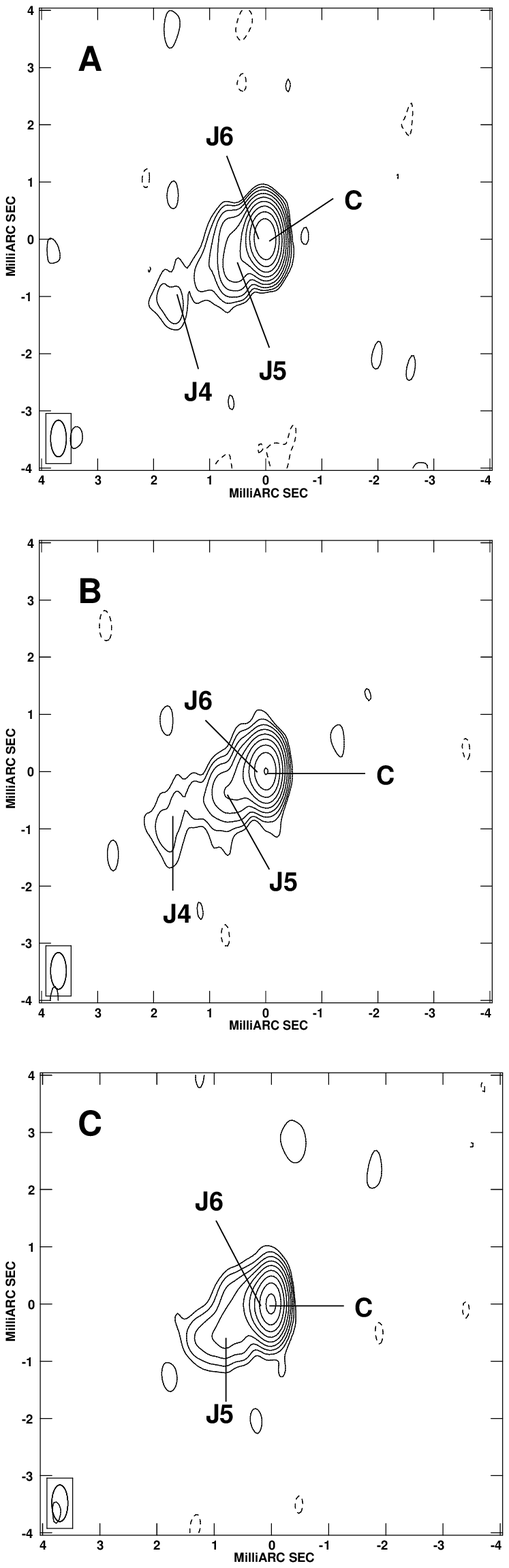}} \vspace{-1mm}
\caption{VLBA images of \object{PKS\, 1502+106} at 24.4 GHz. A:
epoch 2002.37; B: epoch 2002.65; C: epoch 2002.99.} \label{fig3}
\end{figure}

%
%
\begin{figure}\centering \vspace{3mm}
\resizebox{6.5cm}{!}{\includegraphics{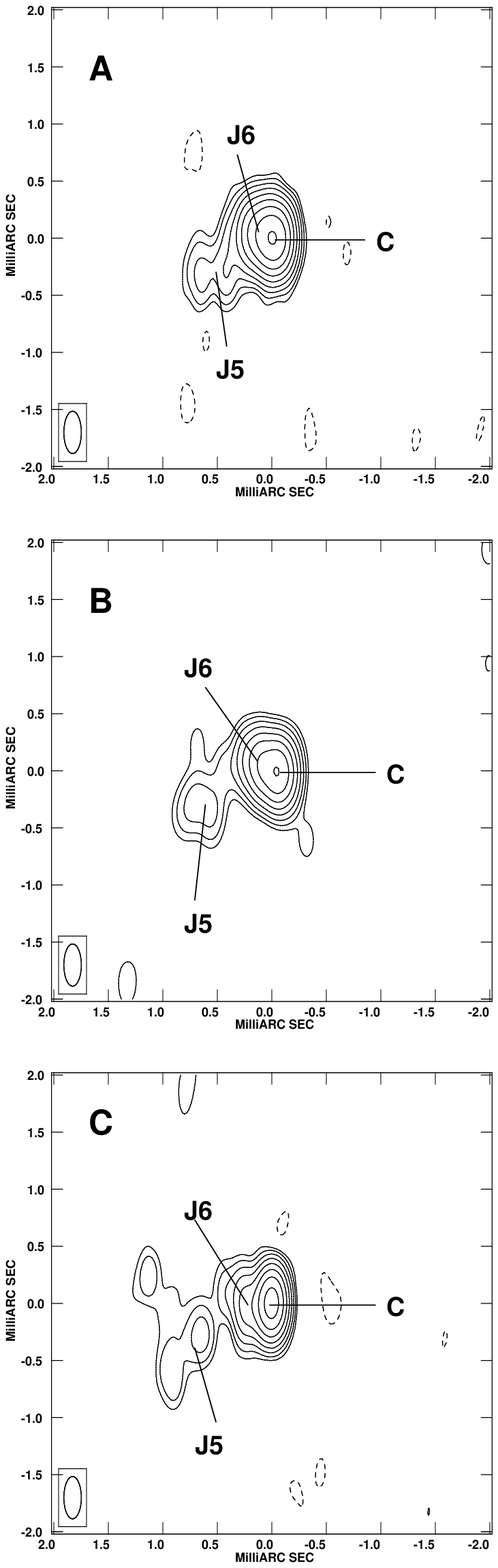}} \vspace{-1mm}
\caption{VLBA images of \object{PKS\, 1502+106} at 43.1 GHz. A:
epoch 2002.37; B: epoch 2002.65; C: epoch 2002.99.} \label{fig4}
\end{figure}

\section{Observations and data processing}

Table \ref{obs} summarizes the observations in order of time. The
epoch and frequency of the observations are given in Cols. 1 and
2. Column 3 gives the bandwidth in each observation. Column 4
presents the array and available antennas. We note that the
observations at epoch 1998.11 were carried out with the VLBA and 4
geodetic antennas. The longest baseline of the array is given in
Col. 5.

\object{PKS\,1502+106} was observed with the EVN at 5 GHz on 7
November 1997. The observation was carried out in snapshot mode,
and \object{PKS\, 1502+106} was observed with 4 scans, each 13
minutes long. The raw data bits were obtained with the
MK\textrm{III} VLBI recording system with an effective bandwidth
of 28~MHz and correlated in Bonn at the Max--Planck--Institute
f\"{u}r Radioastronomie MK\textrm{III} correlator with 4 seconds
integration time. The correlated data bits were calibrated and
corrected for residual delay and rate using NRAO AIPS
(Astronomical Image Processing System). Data post--processing,
including editing, amplitude and phase self--calibration, and
imaging were carried out by means of the AIPS and DIFMAP packages
(Shepherd, Pearson \& Taylor \cite{Shepherd}). The amplitude
uncertainty is about 5\% as estimated from the amplitude gain
calibration. The task MODELFIT in the DIFMAP package was used in
fitting models to the calibrated \emph{uv} datasets.

The five epochs of dual--frequency VLBA archive datasets, observed
from 1994 to 2002, were also re--imaged with natural weight and
analyzed using AIPS and DIFMAP. Details of the observations are
given in Table \ref{obs}. The calibration errors of the VLBA
datasets are mainly caused by receiver noise, and are estimated to
be $\sim$ 1 mJy/beam (Fey, Clegg \& Fomalont \cite{Fey}; Fomalont
et al. \cite{Fomalont}).

\section{Analysis of the images}

The images at 2.3, 5, 8.3, 24.4 and 43.1~GHz are shown in
Figs.~\ref{fig1} to \ref{fig4} in increasing order of frequency.
They are all naturally weighted images. Each figure includes all
epochs at the same frequency. As is clear from Table 2, the rms
noise level ranges from 0.6 to 1.2 mJy beam$^{-1}$. The dynamic
range (ratio of peak intensity to noise level) of the final images
depends on the \emph{uv} coverage and on the quality of the
datasets. Except for the 2002.65 43.1 GHz dataset, whose
\emph{uv}--coverage is rather poor, in all images we obtain high
dynamic ranges, between $\sim$850:1 and $\sim$1700:1. For a proper
comparison, the datasets at 2.3 GHz were re--imaged with the same
cellsize and restored with the same FWHM = $7.18\times3.78$ mas;
the 5 GHz and 8.3 GHz datasets were restored with the same FWHM =
$1.25\times1.25$ mas; the 24.4 and 43.2 GHz images were restored
with a beam of $0.64\times0.28$ mas and $0.37\times0.16$ mas,
respectively. Detailed parameters of the images are given in Table
\ref{image} with the following format: Col. 1 lists the figure
label of the image; cols. 2 and 3 provide the observing epoch and
frequency; col. 4 gives the real beam which is directly obtained
from the self--calibrated visibilities. Column 5 presents the
restored beam shown in each contour image. Columns 6 and 7 give
the peak flux density and the off--source rms fluctuation in the
image, respectively. The rms noise in each image is estimated in
the final CLEANed images. Column 8 lists the contour levels in the
images. The lowest contour level is generally about 3 times the
rms noise level in the image.

\subsection{The parsec--scale morphology}
From the images in Figs. \ref{fig1} to \ref{fig4} it is clear that
\object{PKS\, 1502+106} is characterized by an asymmetric
parsec--scale structure, with a dominant component and a
one--sided jet. We note that images at different frequency and
resolution highlight different features of the parsec--scale
emission. The core region and the position angle of the jet change
considerably over the resolution and frequency range studied here.
Moreover, the images clearly show the morphological variability of
the source with time.

In the 2.3~GHz images (Fig.~\ref{fig1}) the jet points to the east
of the core and extends up to $\sim$~15 mas. No counter--jet
emission is detected in the images. The 5 and 8.3 GHz images in
Fig. \ref{fig2} reveal a curved jet structure within 8 mas. The
inner jet moves out to the southeast of the core, then it turns
sharply to the northeast at a distance of 3--4 mas. No emission
brighter than 4$\times$rms (i.e., 3 mJy/b in Fig. \ref{fig2}a) is
detected beyond 10 mas east of the core. The three--epoch VLBA
images at 24.4 GHz in Fig. \ref{fig3} show the substructure of the
jet in the inner 2 mas. The jet continuously extends to the
southeast, and its position angle is in good agreement with the
inner southeastern jet detected at 5 and 8.3 GHz. Fig. \ref{fig3}a
(epoch 2002.37) and Fig. \ref{fig3}b (epoch 2002.65) show that the
jet brightness decreases beyond 1 mas. The 43.1 GHz images in Fig.
\ref{fig4} make it possible to resolve the core region. The
position angle of the jet within $\sim$ 0.5 mas is $\sim
70^{\circ}$ at epoch 2002.37 and 2002.65 (Figs.~\ref{fig4}a and
\ref{fig4}b respectively), very different from the alignment
beyond 1 mas, where the jet points towards the southeast. We also
note the knotty appearance of the jet at epochs 2002.65 (Fig.
\ref{fig4}b) and 2002.99 (Fig. \ref{fig4}c).

\begin{table*}
\centering{\caption{Component parameters from the model fitting}}

\begin{tabular}{cccccccccc}
\hline
Figure& epoch&Freq & Comp& S  & R   &$\theta$&a    &b/a& p.a.\\
      &      &(GHz)&     &(Jy)&(mas)&(deg)   &(mas)&   &(deg)\\
 (1)  &(2)   &(3)  &(4)  &(5) &(6)  &(7)     &(8)  &(9)&(10)\\
\hline\hline

Fig.1a&1994.52&2.3  &C  & 1.650$\pm$0.276 &0 &0 &0.97 &0.21
&-43 \\
        &       &     &C2 & 0.294           &1.77$\pm$0.37 &118.2 &1.50 & 1
&$\cdots$\\
        &       &     &C1 & 0.165           &6.68$\pm$1.36 &80.0  &5.34 & 1
&$\cdots$\\
        &       &     &S  & 0.023           &8.75$\pm$0.32 &137.2 &1.45 & 1
&$\cdots$\\\hline

Fig.1b&1998.11&2.3  &C  &1.231$\pm$0.160  & 0            & 0 &1.38
&
0.70&-64 \\
        &       &     &C2 &0.141              & 2.62$\pm$0.64&124.0 &2.58 & 1
&$\cdots$\\
        &       &     &C1 &0.143              & 5.67$\pm$1.55&81.0  &6.70 & 1
&$\cdots$\\
        &       &     &S  &0.010            & 9.19         &131.0
&$\cdots$&$\cdots$&$\cdots$ \\\hline

Fig.2a &1997.85& 5    & C   & 0.874$\pm$0.165 & 0            &0 &
0.69 &0.66
&-53     \\
      &       &      & J4  & 0.256           &0.87$\pm$0.21 &132.0 & 0.84 & 1
&$\cdots$\\
      &       &      & J2  & 0.042           &4.00$\pm$0.32 &125.0 & 1.25 & 1
&$\cdots$\\
      &       &      & J1  & 0.060           &6.29$\pm$1.15 &84.0  & 4.60 & 1
&$\cdots$\\
\hline

Fig.2b&1994.52& 8.3 & C  & 1.390$\pm$0.237 & 0            & 0 &
0.45 &0.15
& -42    \\
        &       &     & J3 & 0.218           &0.89$\pm$0.13 &124.5 & 0.50 &1
&$\cdots$\\
        &       &     & J2 & 0.170           &1.94$\pm$0.40 &134.5 & 1.61 &1
&$\cdots$\\
\hline

Fig.2c&1998.11&8.3  &C  & 0.728$\pm$0.191 & 0            &0 &0.58
&0.23
&-72\\
        &       &     &J4 & 0.180           & 0.95$\pm$0.13&116.3 &0.50  & 1
&$\cdots$\\
        &       &     &J3 & 0.031           & 2.29$\pm$0.13&133.6 &$<0.1$& 1
&$\cdots$\\
        &       &     &J2 & 0.027           & 4.39$\pm$0.41&125.8 &1.64  & 1
&$\cdots$\\
        &       &     &J1 & 0.019           & 6.86$\pm$0.32&83.8  &1.28  & 1
&$\cdots$\\
        \hline

Fig.3a&2002.37&24.4 &C  & 1.030$\pm$0.166 & 0              &0
&0.08   &1      &$\cdots$\\
        &       &     &J6 & 0.524$\pm$0.10  & 0.162$\pm$0.028&69.0  &0.11 &1
&$\cdots$\\
        &       &     &J5 & 0.164           & 0.590$\pm$0.128&112.1 &0.51 &1

&$\cdots$\\
        &       &     &J4 & 0.060           & 1.706$\pm$0.330&123.2 &1.32 &1

&$\cdots$\\
        \hline

Fig.3b&2002.65&24.4 &C  & 0.645$\pm$0.115 & 0             &0
&0.13   &1      &$\cdots$\\
        &       &     &J6 & 0.342           & 0.178$\pm$0.043 &82.7  &0.17 &1

 &$\cdots$\\
        &       &     &J5 & 0.129           & 0.713$\pm$0.118 &114.8 &0.47 &1

 &$\cdots$\\
        &       &     &J4 & 0.039           & 1.760$\pm$0.223 &117.1 &0.89 &1

 &$\cdots$\\
\hline

Fig.3c&2002.99&24.4 &C  & 0.870$\pm$0.160 & 0 &0
&0.05   &1      &$\cdots$\\
        &       &     &J6 & 0.251           & 0.190$\pm$0.043 &97.0  &0.17 &1

 &$\cdots$\\
        &       &     &J5 & 0.130           & 0.888$\pm$0.138 &117.6 &0.55 &1

 &$\cdots$\\
\hline

Fig.4a&2002.37&43.1 &C  & 1.050$\pm$0.194 & 0             &0
&0.15 &0.44 &85.3\\
        &       &     &J6 & 0.348           &0.165$\pm$0.040&72.2  &0.16 &1

 &$\cdots$\\
        &       &     &J5 & 0.090           &0.564$\pm$0.100&115.4 &0.40 &1

 &$\cdots$\\
\hline

Fig.4b&2002.65&43.1 &C  & 0.611$\pm$0.197 & 0 &0
&0.14   &1      &$\cdots$\\
        &       &     &J6 & 0.294           &0.186$\pm$0.043&68.4  &0.17 &1

 &$\cdots$\\
        &       &     &J5 & 0.081           &0.770$\pm$0.088&112.0 &0.35 &1

 &$\cdots$\\
        \hline

Fig.4c&2002.99&43.1 &C  & 0.645$\pm$0.252 & 0             &0
&0.07   &1      &$\cdots$\\
        &       &     &J6 & 0.107           &0.211$\pm$0.035&93.7 &0.14 &1

 &$\cdots$\\
        &       &     &J5 & 0.051           &0.750$\pm$0.065&114.0&0.26 &1

 &$\cdots$\\

 \hline \hline
\end{tabular}\label{model}
\end{table*}

\subsection{Analysis of the parsec--scale jet structure}

\begin{figure}\centering
\resizebox{9cm}{!}{\includegraphics{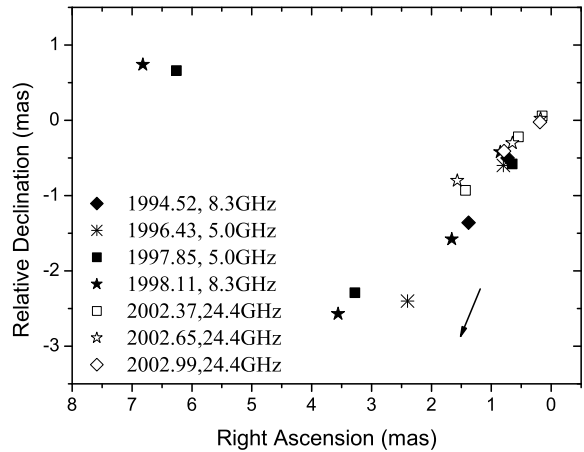}} \vspace{-5mm}
\caption{Two--dimensional projected structure of the VLBI jet in
PKS\,1502+106. Symbols are as follows. \emph{solid diamond}: epoch
1994.52, 8.3 GHz (this paper); \emph{asterisk}: epoch 1996.43, 5
GHz (Fomalont et al. \cite{Fomalont}); \emph{solid square}: epoch
1997.85, 5 GHz (this paper); \emph{solid star}: epoch 1998.11, 8.3
GHz (this paper); \emph{open square}: epoch 2002.37, 24.4 GHz
(this paper); \emph{open star}: epoch 2002.65, 24.4 GHz (this
paper); \emph{open diamond}: epoch 2002.99, 24.4 GHz (this paper).
The arrow indicates the P.A. of the lobe on the arcsecond scale in
the VLA image (Murphy, Browne \& Perley \cite{Murphy}). To avoid
confusion, the component error bars are not shown in the image.}
\label{fig5}
\end{figure}
%
%
\begin{figure}\centering
\resizebox{9cm}{!}{\includegraphics{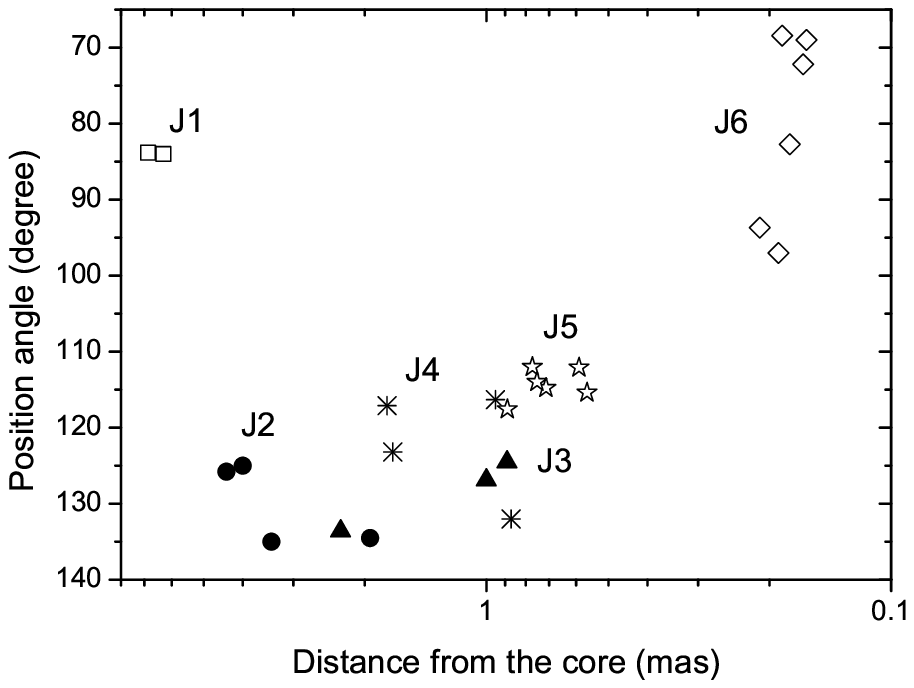}}\vspace{-4mm}
\caption{The distribution of jet component position angles along
the radial distance in PKS\,1502+106. The \emph{x}-axis shows the
distance from the core on a logarithmic scale and \emph{y}-axis
shows the position angle, measured from north to east. Symbols are
as follows: \emph{open square} shows component J1, \emph{solid
circle} J2, \emph{solid triangle} J3, \emph{asterisk} J4,
\emph{open star} J5 and \emph{open diamond} J6.} \label{fig6}
\end{figure}

Model fitting of the radio structure of each image was performed
using the MODELFIT procedure of the Caltech DIFMAP software
package. Table 2 contains the parameters of the major components
for each dataset shown in Figs. \ref{fig1} to \ref{fig4}. Column 1
lists the figure label. Columns 2 and 3 provide the epoch and
frequency of the observation. Column 4 lists the component label.
Component C, assumed to be the stationary core, is the strongest
and most compact component in each image. Due to the different
frequency and resolution, accurate identification of the
components between 2.3 GHz and higher frequencies is impossible.
To avoid possible frequency--dependent misidentifications, we
labelled the jet components at 2.3 GHz as C1, C2 and S, and the
jet components at 5, 8.3, 24.4 and 43.1 GHz with J1 to J6 from the
outermost region going inwards. Column 5 gives the total flux
density of each component. Columns 6 and 7 list the separation and
position angle of each component with respect to the core. Columns
8, 9 and 10 give the parameters of the Gaussian components: the
major axis, the axis ratio and the orientation of the major axis.
An elliptical Gaussian component is usually chosen to fit the
core; however, in those cases where an axial ratio smaller than
0.1 was found, we carried out the fit using a circular Gaussian
component. Circular Gaussians are used for the jet components to
estimate their sizes. The overall uncertainty of the flux density
in each dataset is the quadratic sum of the flux calibration error
and the post--fitted rms error in the image. We present the flux
density errors of discrete, bright components in each dataset. The
errors in the component positions are proportional to the beam
size from the formula given by Fomalont (\cite{Fomalont99}). In
model fitting, we noticed that on average the positions of the
Gaussian components vary less than FWHM/4. For this reason we used
1/4 of the beam major axis as the upper limit of the position
error for most of the components.

At 2.3 GHz we fitted three bright components, labelled as C, C1
and C2 in Fig. \ref{fig1}. Furthermore, at both epochs we found a
weak component (labelled S), located at 8--9 mas to the southeast
of component C.

The source morphology at 5 and 8.3 GHz is well described with four
components along the jet, labelled J1 to J4. Component J3 is
detected at 8.3 GHz at both epochs, but it does not appear at 5
GHz. We failed to fit the extended emission at 3--6 mas in Fig.
\ref{fig2}b with a single gaussian component, due to its extent
and low brightness. We note that component J1 is identified at 5
GHz, epoch 1997.85 (Fig. \ref{fig2}a) and at 8.3 GHz, epoch
1998.11 (Fig. \ref{fig2}c).

Two new components, J5 and J6, are fitted closer to the core at
24.4 and 43.1 GHz (Fig. \ref{fig3} and \ref{fig4}). Component J5
is located at $\sim 0.5 - 1.0$ mas, southeast of the core.
Component J6 is the closest to the core, at $\sim$0.2 mas. We note
the good agreement in the position angle of J5 and J6 at the two
frequencies. Component J4 appears at the end of the southeastern
jet, $\sim$1.7 mas, in P.A.$\sim$120\degr{} (Fig. \ref{fig3}a and
\ref{fig3}b).

Fig. \ref{fig5} displays the projected position of the jet
components in \object{PKS\, 1502+106} at different epochs and
frequencies in an RA--DEC diagram. The symbols represent the
components at different frequencies. Fig.~\ref{fig6} shows the
distribution of jet component position angles with radial distance
from the core. Components J1 to J6 are shown with different
symbols. The investigation of the motions for the individual jet
components suggests a continuous, curved jet path. The jet starts
at a P.A. $\sim70\degr$, and then changes to a P.A. $\sim110\degr$
in the inner 0.5 mas. Another sharp jet bending takes place at
about 3--4 mas from the core, where the position angle changes
from $\sim 130\degr$ to $\sim80\degr$. A helical pattern
associated with hydrodynamical instabilities might explain the
oscillations of the jet position angles (Hardee \cite{Hardee}).

\subsection{Misalignment between the parsec and kiloparsec jet}

\begin{figure}\centering
\resizebox{9cm}{!}{\includegraphics{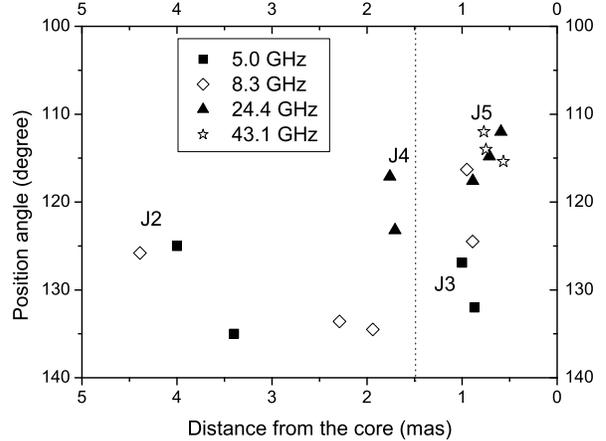}}\vspace{-4mm}
\caption{The frequency-dependence of jet position angles on the
parsec scale in PKS\,1502+106. The \emph{x}-axis and \emph{y}-axis
are defined as radial distance and position angle of the jet
components, respectively. Symbols are as follows. \emph{solid
square}5 GHz, \emph{open diamond} 8.3 GHz, \emph{solid triangle}
24.4 GHz, \emph{open star} at 43.1 GHz. The dashed line represents
the distance at which we measured P.A.$_{pc}$.} \label{fig7}
\end{figure}

To study the radio structure of \object{PKS\, 1502+106}
quantitatively, we examined the jet position angle on different
scales. From the above analysis of the jet structure, the jet
direction oscillates in the range 70\degr--130\degr--80\degr{}
within the inner 8 mas. To have a reference position angle, we
define the P.A. at 10 pc from the core as the value on the parsec
scales (P.A.$_{pc}$). At the distance of \object{PKS\, 1502+106},
10 parsec corresponds to about 1.5 mas (the dashed line in Fig.
\ref{fig7} indicates the distance). We note that the position
angle of the jet in the inner 3 mas changes with frequency (Fig.
\ref{fig7}). In particular, the P.A. at high frequency is
systematically lower than the P.A. at low frequency. The jet axis
is directed to a P.A. of $\sim115\degr$ at 43.1 GHz,
$\sim120\degr$ at 24.4 GHz and $\sim 130\degr$ at 5 and 8.3 GHz.
We believe that this slight discrepancy is due to opacity effects,
and the measurement is frequency dependent. We assume a mean value
of $125\degr$ as the P.A. on parsec scales. The kiloparsec--scale
jet in \object{PKS\, 1502+106}, on the whole, leads to a southeast
lobe along a continuously collimated path (Murphy, Browne \&
Perley \cite{Murphy}). The P.A. of the kiloparsec--scale jet
(P.A.$_{kpc}$) is measured along the direction connecting the peak
and the lobe, and is $\sim157\degr$. Thus the difference of jet
position angles measured on parsec and kiloparsec scales ($\Delta
PA$) is $32\degr$.

%
%

\section{Physical parameters}

\begin{table}\caption{The spectral index of individual component on pc scales}
\begin{tabular}{c|cccccc}\hline
Spectral index& C & J6 & J5 & J4 & J2 & J1
\\\hline
$\alpha^{5}_{8.3}$(1998.11)&-0.36&     &     &-0.70&-0.87&-2.27\\
$\alpha^{24.4}_{43.1}$(2002.37)&0.03 &-0.72&-1.05&\\
$\alpha^{24.4}_{43.1}$(2002.65)&-0.09&-0.27&-0.87\\
$\alpha^{24.4}_{43.1}$(2002.99)&-0.52&-1.64&-1.50\\\hline
\end{tabular}\label{spec}
\end{table}

\subsection{The spectral index in the core region}

Dual--frequency simultaneous VLBA observations allow us to
estimate the spectral index distribution of the source ($S \propto
\nu^{\alpha}$). We used the model fitting results in Table 3 to
estimate the spectral index for the individual components. In
particular, we used the two simultaneous datasets at 24.4 and 43.1
GHz to derive the spectral index of the core and the inner jet
components J5 and J6. Furthermore, we also estimated the spectral
index of components C, J1, J2 and J4 by comparing two
non--simultaneous datasets, 1997.85 5 GHz data and 1998.11 8.3 GHz
data. The results are listed in Table \ref{spec}. As expected, the
spectrum is flat in the core region, and steeper along the inner 4
mas jet. We note that the spectrum is systematically steeper at
epoch 2002.99, and the flux density of each jet component
decreases from epoch 2002.65 to epoch 2002.99 both at 24.4 and
43.1 GHz. This could be due to an expansion of the jet. The 5 GHz
EVN data and 8.4 GHz VLBA data are not simultaneous and the
\emph{uv} coverage of the two arrays is also different, therefore
the spectral index estimated between these two frequencies should
be taken with caution. The total flux density decreased by 21\%
from epoch 1997.85 to epoch 1998.11.

\begin{table*}
\caption{Intrinsic Parameters and Doppler Factor in the radio
core}
\begin{tabular}{c|ccccccccc}\hline

Epoch &1997.85&1994.52 &1998.11 &2002.37&2002.65
&2002.99&2002.37&2002.65&2002.99\\\hline

Freq(GHz) &5&8.3 &8.3 &24.4 &24.4 &24.4 &43.1 &43.1 &43.1\\\hline

T$_{b}^{(a)}$($10^{12}$K) &0.38 &2.30 &0.47 &0.93&0.22 &2.02 &0.20
&0.06&0.24\\\hline

$\delta_{eq}$ &1.87&15.0 &2.45 & 5.20 &0.99 &13.0&0.91 &0.21 &1.14
\\\hline

T$_r^{(b)}$ ($10^{12}$K) &0.20 &0.15 & 0.19 &0.18 &0.22 & 0.16
&0.22& 0.27&0.21
\\\hline

\end{tabular}\vspace{3mm}
\label{TbTable}\\
\raggedright $^{(a)}$ observed radio core brightness temperature;\\
$^{(b)}$ intrinsic radio core brightness temperature, taking into
account the Doppler boosting factor.
\end{table*}

\subsection{Equipartition Doppler factor and brightness Temperature }

Assuming that the source is at or near equipartition of energy
between radiating particles and magnetic field, we can estimate
the equipartition Doppler factor $\delta_{eq}$ of the radio core
(Readhead \cite{Readhead} and G\"{u}ijosa \& Daly \cite{Guijosa})
with the formula:

\begin{eqnarray}
\delta_{eq}=[[10^3 F(\alpha)]^{34}\{[1-(1+z)^{-1/2}]/2h\}^{-2}
S^{16}_{op}\theta^{-34}_d\nonumber \\
\times (1+z)^{15-2\alpha} (\nu_{op}\times
10^3)^{-(2\alpha+35)}]^{1/(13-2\alpha)}, \label{delta}
\end{eqnarray}

where \emph{z} is the redshift and $\theta_{d}$ is the angular
diameter of the source. In computing the Doppler factors, we
assumed an optically thin spherical index $\alpha=-0.75$ and a
spherical geometry, as was used by G\"{u}ijosa \& Daly
(\cite{Guijosa}). $h=0.65$ is defined. For an elliptical gaussian
component, $\theta_{d}$ is larger than the observed angular
diameter $\theta_{FWHM}$. Marscher (\cite{Marscher}) suggested a
correction for this by using $\theta_{d} = 1.8\theta_{FWHM}$.
$S_{op}$ is defined as the observed flux density at the
self-absorption turnover frequency $\nu_{op}$.

The equipartition method requires that observations are carried
out at the self--absorption turnover frequency. However it is
difficult to obtain the VLBI core parameters in PKS 1502+106 at
the exact turnover frequency, since the core shows a flat spectrum
between the observing frequencies from 2.3 to 43.1 GHz (Table
\ref{spec}). Assuming the VLBI observing frequency as the turnover
frequency, we roughly estimate the Doppler boosting factor using
the available VLBI core parameters. The formula for the
equipartition Doppler factor is geometry dependent. The Doppler
factor in a jet model is slightly larger than that in the
spherical case for $\delta >1$,
$\delta_{eq}(jet)\propto\delta_{eq}^{1.4}(sph)$. We note that the
most compact component at lower frequencies is resolved into a
compact component and initial jet at higher frequencies.
Particularly, the core component at 2.3 GHz includes the core and
the southeast innermost jet portion. So we limited the estimates
of the Doppler factors to the data of the most compact component C
at 5, 8.3, 24.4 and 43.1 GHz to minimize the geometry--dependence
uncertainty. The estimated $\delta_{eq}$ values are listed in
Table \ref{delta}. We assume a mean value $\delta=4.5\pm1.9$ as
the best estimate of the Doppler boosting factor in the core of
PKS 1502+106. In practice, a low limit to $\delta_{eq}$ in PKS
1502+106 has been calculated with an assumed optically thin
spectral index ($\alpha=-0.75$) in the formula (Readhead
\cite{Readhead}).

The brightness temperature $T_b$ of a Gaussian component in the
source rest frame is given by Shen et al. (\cite{Shen}):
\begin{equation}
T_b = 1.22 \times 10^{12}
\frac{S_{op}}{\nu^{2}_{op}ab}(1+z)\;\;\;K \label{Tb},
\end{equation}

where a and b are the major and minor axes in unit of mas, other
parameters are defined same as Eq. (1). The radio core flux
density and dimensions are taken from Table 3. Taking into account
the Doppler boosting factor, we get the equivalent value of the
intrinsic brightness temperature $T_{r}$ ($T_{r} = T_{b}
/{\delta}$). The estimated values of $T_b$ and $T_r$ are presented
in Table \ref{TbTable} and we assume a mean value $T_r= (2.0\pm
0.5)\times10^{11} K$ as the best estimate of the intrinsic
brightness temperature in the source frame. The major errors of
$T_r$ are introduced by the fitted core size and the source
variation. Due to the resolution limit, we only obtain the lower
limit of the brightness temperature. Readhead (\cite{Readhead})
suggested that the equipartition temperature is likely to be
$10^{10}\sim10^{11}$K in most sources. We can see that the
intrinsic radio core brightness temperature in the source,
($2.0\pm0.5)\times 10^{11}$K, approaches the equipartition limit.

\subsection{Superluminal motion in the parsec--scale jet}

\begin{figure}
\resizebox{9cm}{!}{\includegraphics{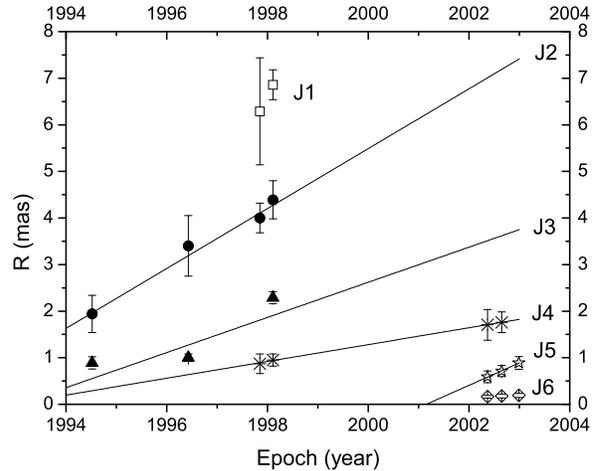}} \vspace{-8mm}
\caption{Distances of the jet components from the core as a
function of time. We included the data point at 5 GHz at epoch
1996.43 taken from Fomalont et al. (\cite{Fomalont}). All the
other epochs are those presented in this paper. The line
represents the linear fitting of the motion for each jet
component.} \label{fig8}
\end{figure}

In the discussion of the parsec--scale jet structure (Section
3.2), we suggested that the jet components in \object{PKS\,
1502+106} follow the same curved trajectory. For component J6 we
find only a marginal indication of radial motion, however we note
that its position angle changes significantly at the same radial
distance from the core. The extended component J1 is detected only
in two epochs, therefore its proper motion is estimated with a
large uncertainty. For these reasons, we only calculate and
discuss the proper motion for the well--defined components J2, J3,
J4 and J5. The time evolution of the position of each component
with respect to the (assumed) stationary core is reported in
Fig.~\ref{fig8}. We underline that due to possible opacity effects
and frequency--dependent offsets in the position, only the
Gaussian components at the three frequencies 5, 8.3 and 24.4 GHz
are considered in the study of the motion. We also include the 5
GHz Gaussian models presented by Fomalont et al. (\cite{Fomalont})
for a more complete analysis (see caption to Fig.~\ref{fig8}).

We performed least--squares fits to the component distances from
the cores as a function of time. To minimize the effects of
frequency--dependent separation we took the model fitting errors
as the statistical weight and measured the overall weighted mean
velocities. The results are presented in Table \ref{PM}, where
$\mu$ represents the proper motion and $\beta_{app}$ denotes the
apparent velocity. The measured proper motions range between
$0.18\pm0.04$ and $0.64\pm0.16$ mas/yr, corresponding to apparent
velocities in the range between $10.5\pm2.6\,c$ and
$37.3\pm9.3\,c$. These estimated speeds are significantly higher
than the average superluminal speed of jet components in radio
loud quasars (Pearson et al. \cite{Pearson}). Jorstad et al.
(\cite{Jorstad}) found that the mean apparent velocities of
$\gamma$-ray quasars increase with redshift. The average velocity
for high redshift ($2\leq z \leq 2.5$) quasars is $15.9\pm6.6\,
h^{-1}c$ (i.e., $24.5\pm10.2\,c$ for $h=0.65$). Except for the
highest superluminal velocity, $37.3\pm9.3\,c$, the apparent
speeds in PKS\,1502+106 ($z=1.833$) are in good agreement with the
statistical mean value of Jorstad et al. (\cite{Jorstad}).

We note that the highest speed is associated with the jet
component J2, located at the position of the jet bending, i.e.
3--4 mas from the peak. A possible explanation for the very high
speed of J2 is that its observed properties are magnified by
projection effects. We note that at the very high levels of
beaming revealed in this source, a deflection of only $\sim
1^{\circ}$ is enough to increase the apparent speed of component
J3 to that of component J2. In particular, from the theory of
superluminal motion (Pearson \& Zensus \cite{Pearson87}) the
minimum viewing angle $\theta_{min}$ goes from $2.60^{\circ}$ to
$1.54^{\circ}$ if the apparent speed increases from $\beta_{app} =
22.0 c$ (component J3) to $37.3 c$ (component J2).

\begin{table}
  \centering
  \caption{Proper motion of jet components in PKS\,1502+106}
  \begin{tabular}{c|c|c}\hline
   Comp & $\mu$ (mas/yr)& $\beta_{app}$ \\\hline
   J2   &$0.64\pm0.16$  &$37.3\pm9.3\,c$\\
   J3   &$0.38\pm0.26$  &$22.0\pm15.5\,c$\\
   J4   &$0.18\pm0.04$  &$10.5\pm2.6\,c$\\
   J5   &$0.48\pm0.12$  &$27.9\pm7.0\,c$\\\hline
  \end{tabular}\label{PM}
\end{table}

%
%

\section{Conclusions and Summary}

In this paper we carried out a multifrequency and multiepoch study
of \object{PKS\, 1502+106} at VLBI resolution. The source
morphology is highly variable and the jet structure is very
complex on this scale.

We analyzed the structure of the jet, as seen in projection on the
sky, and the changes in the position angle of the various jet
components with the radial distance from the core. The results
suggest that the jet components in \object{PKS\, 1502+106} trace
the same curved path. The jet undergoes two major bends, the first
implies a change in the P.A from $\sim70\degr$ to $\sim110\degr$
within 0.5 mas, while in the second bend the P.A. changes from
$\sim130\degr$ to P.A.$\sim80\degr$ at 3--4 mas. Beyond that
distance the VLBI jet points to the east at 7--15 mas.

Based on the model fitting results on the VLBI core, we obtain a
weighted mean value of the Doppler factor, $\delta=4.5\pm1.9$. The
radio core brightness temperature in the source rest frame, T$_r =
(2.0\pm0.5) \times10^{11}$K, approaches the equipartition limit.

We detect superluminal motion in four components along the jet.
The derived apparent speeds range between $10.5\pm2.6\,c$ and
$37.3\pm9.3\,c$. Doppler boosting plays a major role in
determining the observed properties of the source. In particular,
the apparent speeds we derive suggest that the source is viewed
under an angle to the line of sight $\theta < 5^{\circ}$, and that
the bulk flow velocity is $\beta_{intr}$ is $\sim$ $0.999\,c$.

The $\Delta PA$ between the pc-- and kpc--scale structure is about
30\degr{}, indicating that PKS\,1502+106 belongs to the aligned
population. The superluminal speeds in 1502+106 are much higher
than the average value in radio loud quasars. We therefore
conclude that PKS\,1502+106 is more beamed than the overall
population of radio loud quasars, and that its radio properties
are more similar to the $\gamma$--ray loud quasars, although it is
unclear if \object{PKS\, 1502+106} is a $\gamma$--ray loud source
to date. In particular, the superluminal speeds in this source are
in the range found for the high redshift $\gamma$--ray loud
quasars in other high frequency radio surveys (Jorstad et al.
2001).

A confirmation of $\gamma$--ray emission from this source would be
highly valuable for our understanding of the $\gamma$--ray
loudness phenomenon in radio loud quasars.

\begin{acknowledgements}

This research was supported by the National Science Foundation of
PR China (10333020, 10328306 and 10373019), Foundation of Chinese
Academy of Sciences (19973103). T. An thanks IRA of CNR for their
hospitality during his visit to Italy for the data reduction in
2001. The studies reported in this paper have been supported by
the grant for collaborative research in radio astronomy of the CAS
(Chinese Academies of Sciences) and CNR (the Italian National
Research Council), Posiz. N. 132.20.1. This research has made use
of the United States Naval Observatory (USNO) Radio Reference
Frame Image Database (RRFID). The authors thank A.L. Fey for
providing access to the RRFID datasets. The authors are also
grateful to the staff of all EVN radio observatories and data
reduction center, as well as the Hartebeesthoek observatory for
support during the observations. The European VLBI Network is a
joint facility of European, Chinese, South African and other radio
astronomy institutes funded by their national research councils.
This work has made use of NASA Astrophysics Data System Abstract
Service.
\end{acknowledgements}

\end{document}